\documentstyle[12pt,ssi,epsfig]{article}

\begin{document}

\def\sintw{\sin^2 \theta_W^{\overline{MS}} (M_Z)}
\def\sintwQ2{\sin^2 \theta_W^{\overline{MS}} (Q^2=0.026 {\rm GeV}^2)}
\def\sintwMZ{\sin^2 \theta_W^{\overline{MS}} (M_Z^2)}

\def\thetaweff{\sin^2 \theta_W^{eff}}
\def\msbar{\overline{MS}}
\def\alr{A_{LR}}
\def\apv{A_{PV}}
\def\apvmoller{A_{PV}^{{\rm M}\o{\rm ller}}}
\def\moller{{\rm M}\o {\rm ller}}

\def\z0{Z^0}
\def\rar{\rightarrow}
\def\beamalr{^{\rm beam}{\rm A}_{\rm LR}}
\def\bkgdalr{^{\rm bkgd}{\rm A}_{\rm LR}}
\def\apvphys{A_{PV}^{phys}}
\def\apvmeas{A_{PV}^{meas}}
\def\pole{P_e}

\begin{flushright}
{\small
SLAC--PUB--10338\\
February, 2004\\
Revision 1:  May, 2004}
\end{flushright}

\title{A New Measurement of the \\ Weak Mixing Angle}

\author{Mike Woods,\thanks{Supported by the Department of Energy, 
Contract DE-AC03-76SF00515.} \\ 
Stanford Linear Accelerator Center \\
Stanford University, Stanford, CA 94309 \\[0.2cm]
Representing the SLAC E158 Collaboration
}

\maketitle
\begin{abstract} The E158 experiment at SLAC has made the first measurement of parity violation in electron-electron ($\moller$) scattering.  We report a preliminary result using 50\% of the accumulated data sample for the right-left parity-violating cross-section asymmetry ($\apv$) in the elastic scattering of 45 and 48 GeV polarized electron beams with unpolarized electrons in a liquid hydrogen target. We find $\apv = (-160 \pm 21 {\rm (stat.)} \pm 17 {\rm (syst.)}) \cdot 10^{-9}$, with a significance of 6.3$\sigma$ for observing parity violation.  In the context of the Standard Model, this yields a measurement of the weak mixing angle, $\sintwQ2 = 0.2379 \pm 0.0016 {\rm (stat.)} \pm 0.0013 {\rm (syst.)}$.  We also present preliminary results for the first observation of a single-spin transverse asymmetry in $\moller$ scattering.
\begin{center}
{\small 
\vskip 4mm
Presented at \\
\vspace{3mm}
{\it $31^{st}$ SLAC Summer Institute on Particle Physics:  \\
Cosmic Connections to Particle Physics (SSI 2003)  \\
July 28--August 8, 2003; Menlo Park, CA}}  \\
\end{center}

\end{abstract}


\pagebreak

\section*{1.  Introduction and Physics Motivation}
The weak mixing angle has been precisely measured at the Z-pole by the LEP experiments at CERN and by the SLD experiment at SLAC, from measurements of left-right and forward-backward asymmetries and from tau polarization.\cite{sldlep}  But additional precise measurements away from the Z-pole are needed to probe for certain classes of new physics, and to test the Standard Model predictions for the running of $\sin^2 \theta_W$ with $Q^2$.\cite{marciano2}  Interaction amplitudes measured at the Z-pole are almost purely imaginary and may have negligible interference with interaction amplitudes involving heavy new particles.  Such interference effects may show up more readily with measurements at low $Q^2$ away from the Z-pole.~\cite{proposal,kk}

\begin{figure}
\centering
\epsfig{file=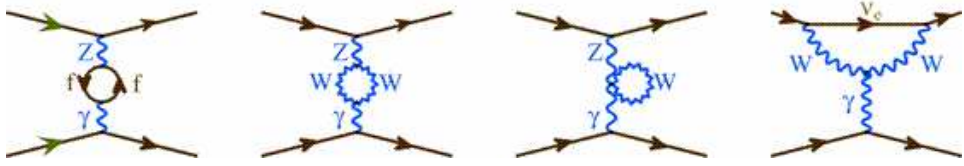,width=5.0in}
\caption{ Some of the radiative corrections that cause the $Q^2$-dependence of the weak mixing angle.}
\label{fig:RCs}
\end{figure}

E158 measures the parity-violating cross-section asymmetry 
\begin{equation}
\apv = \frac{(\sigma_R - \sigma_L)}{(\sigma_R + \sigma_L)}
\end{equation}
for small angle $\moller$ scattering with very high statistics,\cite{proposal} where $\sigma_R$ and $\sigma_L$ are the scattering cross sections for incident right- and left-polarized beams.  This $\apv$ arises from interference between the weak and electromagnetic amplitudes\cite{zeldovich} and is sensitive to the weak mixing angle.  The tree-level expression for $\apv$ is given by\cite{marciano1} 
\begin{equation}
A_{PV} = \frac{G_FQ^2}{\sqrt{2} \pi \alpha} \cdot \frac{1-y}{1+y^4+(1-y)^4} \cdot (1-4\sin^2 \theta_W), 
\label{eq:apvtree}
\end{equation}
where $G_F$ is the Fermi constant; $Q^2$ is the momentum transfer and is $\approx 0.03$ GeV$^2$ for the E158 kinematics; $\alpha$ is the fine structure constant; and $y=Q^2/s$ ($<y> \approx 0.6$ for E158).  The expected asymmetry at tree level is approximately $3.2 \cdot 10^{-7}$.\cite{marciano1}  Radiative corrections (Figure~\ref{fig:RCs}) reduce this asymmetry by about $50\%$.~\cite{marciano2,marciano3,erler,petriello,ferroglia}  

\begin{figure}
\centering
\epsfig{file=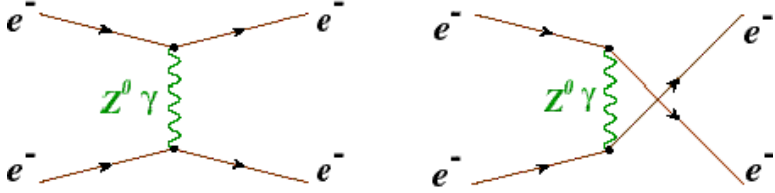,width=4.0in} 
\caption{ Tree-level Feynman graphs for electron-electron ($\moller$) scattering.}
\label{fig:feynman}
\end{figure}

E158 continues a successful program at SLAC studying parity violation (PV) in the weak neutral force that began with Charles Prescott$'$s historic experiment E122 in 1978.\cite{prescott}  E122 made the first observation of parity violation (PV) in a weak neutral scattering interaction by observing PV in deep inelastic scattering of polarized 19.4 GeV electrons from an unpolarized liquid deuterium target.  It was one of the cornerstone experiments that solidified the Standard Model (SM) developed by Glashow, Weinberg and Salam to describe electroweak interactions.  

In the LEP and SLC era in the 1990s, precision measurements probing quantum effects from physics at higher energy scales were very successful.  Precision electroweak measurements accurately predicted the mass of the top quark before it was discovered at the Tevatron,\cite{top} and they were cited in the awarding of the 1999 Nobel Prize to Veltmann and t$'$Hooft,\cite{nobel} which recognized their work in developing powerful mathematical tools for calculating quantum corrections and demonstrating that the SM was a renormalizable theory.

\begin{figure}
\centering
\epsfig{file=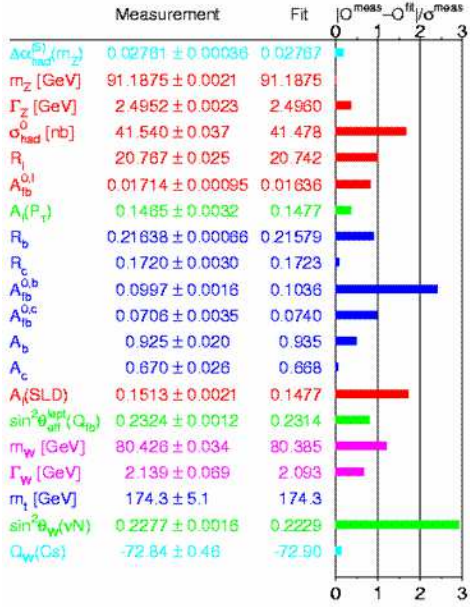,width=2.5in}
\caption{ A summary~\cite{sldlep} of precision electroweak measurements at LEP, SLC and the Tevatron; also from the fixed target neutrino-nucleon scattering experiment, NuTeV,~\cite{nutev} and the Cs atomic parity violation experiment.~\cite{cesium}  The graph on the right indicates the $"$pulls$"$ of each measurement; ie. the number of standard deviations from the theoretical fit values.}
\label{fig:ewsummary}
\end{figure}

\begin{figure}
\centering
\epsfig{file=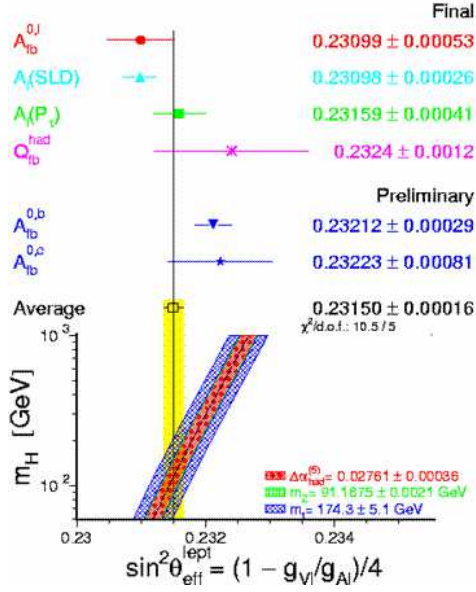,width=2.5in}
\caption{ A summary~\cite{sldlep} of (Z-pole) weak mixing angle measurements at LEP and SLC.}
\label{fig:weakanglelepsld}
\end{figure}

The discovery and mass measurement of the top quark at the Tevatron and the precise Z boson mass measurement from LEP complement well established values for $G_F$ and $\alpha$, and allow predictions for the only SM parameter not yet measured, the Higgs mass, from other electroweak observables.  A recent compilation of electroweak measurements and their consistency is presented in Figure~\ref{fig:ewsummary}.  A subset of these measurements, using forward-backward asymmetry and tau polarization measurements at LEP and the left-right asymmetry measurement at SLC, give precise measurements of the weak mixing angle at $Q^2=M_Z^2$.  These Z-pole weak mixing angle results are presented in Figure~\ref{fig:weakanglelepsld}.  

The data presented in Figure~\ref{fig:ewsummary} have a large $\chi^2/dof = 27.3/15$, with a probability of only 3$\%$.  The weak mixing angle data in Figure~\ref{fig:weakanglelepsld} also have a large $\chi^2$ and indicate an inconsistency between the SM leptonic and hadronic couplings.  Perhaps we are starting to observe in these data some cracks in the Standard Model.  And perhaps there are bigger effects in precision electroweak measurements away from the Z-pole.  Indeed the largest discrepancy in Figure~\ref{fig:ewsummary} is from a measurement of the weak mixing angle by the NuTeV experiment,\cite{nutev} using the observed relative rates of charged and neutral current events in neutrino-nucleon scattering.  The NuTeV result is 3 standard deviations from the SM prediction at a $Q^2 \approx 30$ GeV$^2$ and possibly results from new physics at a higher energy scale, such as a heavy Z$'$.  New results from E158 can add important information to these interesting deviations from SM predictions that are observed in the precision electroweak observables.

E158 performed 3 physics runs in 2002 and 2003.  We report here on the final result from Run I~\cite{run1prl} and a preliminary result from Run II, which together comprise about 50$\%$ of the entire data sample.  More details on the experiment, apparatus and analysis can be found in the Ph.D.
thesis by David Relyea.~\cite{relyea}

\vspace{8.5mm}
\section*{2.  Polarized Beam and Beam Monitors}

The electron beam is produced by photoemission, using a circularly polarized laser beam~\cite{humensky} and a strained GaAs photocathode.~\cite{maruyama}  The electron polarization is $85\%$.  The beam~\cite{turner,decker} is accelerated to high energy in the two-mile SLAC Linac and then transported through a $24.5^{\circ}$ bend angle in the A-line~\cite{erickson} to a liquid hydrogen (LH$_2$) target in End Station A(ESA).  Pulsed magnets were installed in the DRIP (Damping Ring Intersection Point) region of the Linac to allow interleaved operation of the E158 experiment with the BaBar experiment at the PEP-II storage ring.~\cite{decker}  A skew quad was also installed at the end of the A-Line for E158 to take advantage of the horizontal emittance growth in the A-Line.\cite{woodley}  The skew quad provides {\it x-y} coupling and allows a more stable beam spot with less spatial tails at the LH$_2$ target.

\begin{figure}
\centering
\epsfig{file=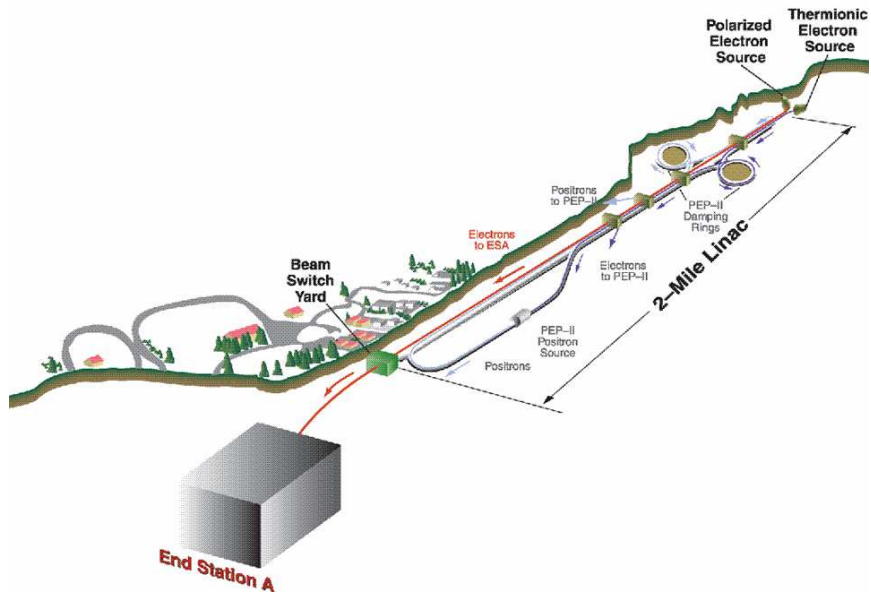,width=4.5in}
\caption{ Schematic of the Polarized Source, Linac and A-Line beam transport to E158 in End Station A.}
\label{fig:linac}
\end{figure}

The electron beam spin is longitudinal at the source and remains longitudinal in the Linac.  In the A-line bend magnets, the spin precesses $180^{\circ}$ every 3.2 GeV.  The spin is longitudinal at the target for a beam energy of 45.6 GeV and 48.7 GeV.  A schematic of the polarized source, Linac and A-Line beam transport is shown in Figure~\ref{fig:linac}.

\begin{table} [tbp]
\caption{Summary of delivered beam parameters for E158, and comparison with projected parameters for  the Next Linear Collider operating at 500 GeV center-of-mass energy.~\cite{turner}} 
\begin{center}
\begin{tabular}{|c|c|c|}
\hline
{\bf Parameter}	& {\bf E158} & {\bf NLC-500}	\\
\hline
Charge/Train 			& $5 \cdot 10^{11}$	& $14.4 \cdot 10^{11}$	\\
Repetition Rate		& 120 Hz		& 120 Hz 			\\
Energy				& 45 GeV		&  250 GeV			\\
e$^-$ Polarization		& 85$\%$ 		&  85$\%$			 \\
Train Length			&  270 ns		&  267 ns			 \\
Microbunch spacing		& 0.3 ns		&  1.4 ns 			\\
Beam Loading			& 13$\%$		&  22$\%$ 			\\
Energy Spread			& 0.15$\%$		&  0.3$\%$ 			\\
\hline
\end{tabular}
\end{center}
\label{tab:beamparams}
\end{table}

\begin{figure}
\centering
\epsfig{file=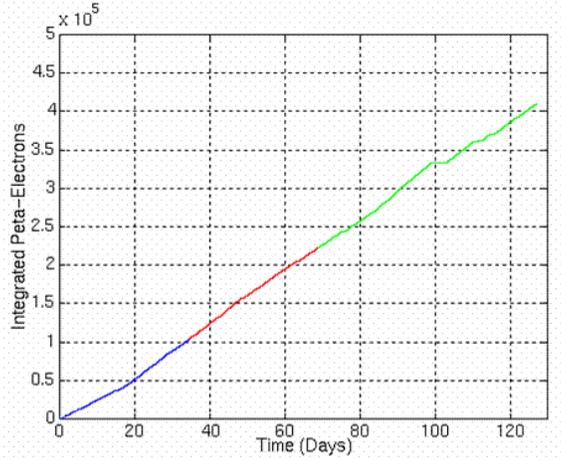,width=3.0in}
\caption{ Beam delivery for E158, indicating the total number of delivered electrons in the 3 physics Runs.  Run I is indicated in blue, Run II in red and Run III in green.  One Peta-Electron is $10^{15}$ electrons.  A total of $4.1 \cdot 10^{20}$ electrons were delivered to the experiment.}
\label{fig:petaE}
\end{figure}

The beam is pulsed at 120 Hz with an intensity of $5 \cdot 10^{11}$ electrons in a 300ns spill. The relevant beam parameters are shown in Table~\ref{tab:beamparams}, and are compared to the expected beam parameters at a 500 GeV Linear Collider.  The pulse charge for the E158 beam approaches what is required for NLC.  (The polarized source can actually produce up to $30 \cdot 10^{11}$ electrons in 270 ns, but the SLAC Linac cannot accelerate that much charge while maintaining a low energy spread.)  It is a factor 15 higher than that used for SLC operation and has a pulse structure similar to that proposed for NLC.  The E158 beam power can exceed 0.5 MW and achieves good stability.  The beam delivery efficiency for E158 Run I was measured to be 65$\%$ (comparing delivered pulses to a continuous request of 120Hz pulses), including the inefficiency due to sharing pulses with PEP.  E158 Run II was a dedicated run with no pulse sharing with PEP and the beam delivery efficiency achieved was 75$\%$.  The E158 beam performance, including efficiency, met or exceeded all of the beam design goals.\cite{proposal}  A summary of the beam delivery for E158 is shown in Figure~\ref{fig:petaE}. 
 
Careful measurements of the beam parameters are required to determine contributions of any false beam asymmetries, $\beamalr$$'$s, to the measured physics asymmetries.~\cite{maestro}  This is discussed in more detail in the following sections.  Precision rf-cavity beam position monitors (BPMs) are used to measure the position and angle of the {\it left, right} beams on the E158 target.~\cite{yury}  Two redundant {\it position} BPMs with $\approx 2$ micron per spill resolution are located 2 meters upstream of the target. An additional pair of {\it angle} BPMs are located 40 meters further upstream and also have $\approx 2$ micron per spill resolution.  Partway through the A-Line between the Linac and ESA are two {\it energy} BPMs at points of high dispersion ($\eta_x \approx 500$ mm) and on opposite sides of the momentum-defining slits (2$\%$ full energy aperture).  The energy resolution of these BPMs is $\approx 1$ MeV per spill.  For measuring the beam spotsize at the LH$_2$ target, there is an {\it x-y} wire array with 300 micron pitch just upstream of the LH$_2$ target.  Additionally, there are 3 BPMs and 2 toroids located at the 1 GeV region of the accelerator for use in the polarized source helicity feedbacks, as described in the next section.

For E158 we added a Synchrotron Light Monitor (SLM) detector in the A-Line to measure (L-R) asymmetries in the emitted synchrotron radiation (SR) that can result from a vertical component of the beam polarization in the horizontal A-Line bend magnets (with vertical B-field).\cite{bondar,belo}  This detector is located behind an existing SR diagnostic port.  It consists of a 1-mm thick lead pre-radiator downstream of an aluminum vacuum flange and a quartz bar to generate Cherenkov light from Compton electrons and pairs, primarily produced by SR photons above 1 MeV energy.  The Cherenkov light is channeled by aluminized mirrors and an air light guide to three one-inch diameter photodiodes.  For SR with energies larger than the critical energy, the SR asymmetry can be approximated by $A(SR) \approx \frac{E_{photon}}{E_{beam}}$.\cite{belo}  Hence, 5 MeV SR photons have an asymmetry of $\approx 10^{-4}$.  An SR asymmetry can generate an energy asymmetry in the beams (though this is measured with the energy BPMs) and could also generate a background asymmetry in the MOLLER detector, if there is a sizable SR background.  We do observe a non-zero SR asymmetry in the SLM detector, indicating $P_y \approx (1-2)\%$.  The dependence of the SLM asymmetry with energy and half-waveplate configurations is consistent with the effect being due to a (roughly) static vertical misalignment in the Linac or A-Line, and there is good cancellation from the two energy configurations for the net effect on the Moller asymmetry.

\begin{figure}
\centering
\epsfig{file=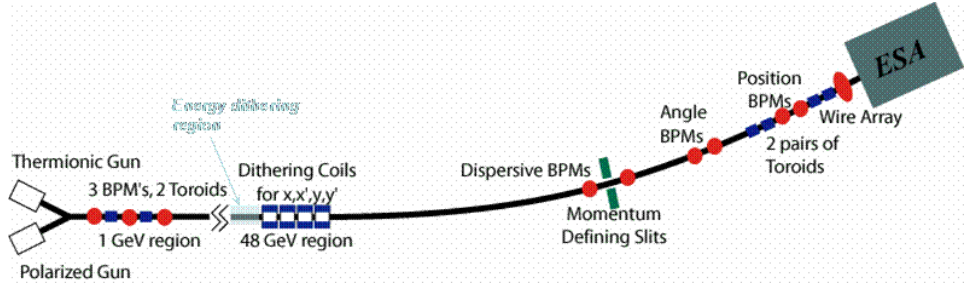,width=5.0in}
\caption{ Beam Monitors in the Linac and A-Line.}
\label{fig:beammonitors}
\end{figure}

\section*{3.  {\it Left-Right} Beam Asymmetries: $\beamalr$$'$s}

While the SLD experiment measured a large $\approx 12\%$ asymmetry for $\alr$ in the production of $\z0$ particles in electron-positron annihilation,~\cite{sldalr} the E158 experiment seeks to measure a small $\approx 160$ part-per-billion (ppb) asymmetry arising from the interference of $\z0$ and photon exchange in electron-electron scattering.  Measuring an asymmetry this small poses 
many experimental challenges!  One challenge is to provide a high intensity beam with negligible left-right asymmetries in the beam parameters (intensity, energy, position, angle), $\beamalr$$'$s.  We want to be sure we are measuring the 160 ppb physics asymmetry in $\moller$ scattering, and not the $\beamalr$$'$s.  For example, we measure a very high scattered electron flux ($\approx 10^7$ scattered electrons per pulse) at small scattering angles ($\approx 5$ mrad) and are very sensitive to left-right targeting differences of the electron beam on the LH$_2$ target.  The electron beam spotsize at the target is roughly 1mm in size, yet a 10 nanometer left-right position difference averaged over the experiment could yield a false asymmetry at the level of 10 ppb! 

The measured asymmetry in the MOLLER detector may be written as
\begin{equation}
\apvmeas=\apvphys+\beamalr^{Det}+\bkgdalr 
\label{eq:beamalr}
\end{equation}
where $\apvphys$ is the $\moller$ physics asymmetry to be determined, and $\beamalr^{Det}$ and $\bkgdalr$ are contributions to the measured asymmetry from beam asymmetries and from backgrounds.  Background asymmetry contributions are discussed in Section 5 below.  The contributions to $\beamalr^{Det}$ arise from (L-R) beam differences, $\beamalr$$'$s, in beam charge, energy, position and angle at the LH$_2$ target in ESA.  The $\beamalr$$'$s originate in the polarized source and a number of techniques are used to minimize them.~\cite{humensky}

\begin{figure}
\centering
\epsfig{file=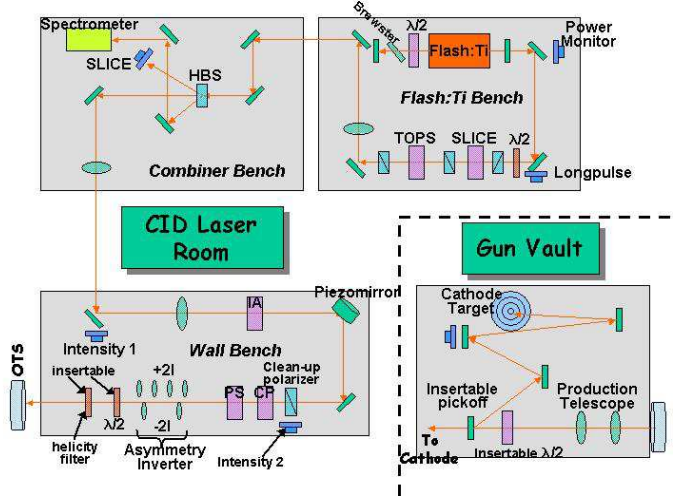,width=3.5in}
\caption{ Optics configuration for the Polarized Source Laser System.  The E158 DAQ controls the IA, CP and PS Pockels cells and the piezo mirror, all located on the Wall Bench.  The DAQ also reads out the position status of the half-waveplate, helicity filter and {\it Asymmetry Inverter} optics.}
\label{fig:helconben}
\end{figure}

The source laser optics (CP and PS Pockels cells, piezo mirror, and IA Pockels cell) to control the laser polarization, {\it left-right} laser steering differences and {\it left-right} charge asymmetries are shown in Figure~\ref{fig:helconben}.  These devices are all controlled by the E158 Data Acquisition (DAQ) and can have different control voltages for {\it left} and {\it right} beam pulses.  The laser is circularly polarized using a linear polarizer and the CP and PS Pockels cells, as shown in Figure~\ref{fig:helconben}.  This configuration can generate arbitrary elliptical polarization, and can compensate for phase shifts in the laser transport optics. The electric field vector following the PS cell can be expressed by 
\begin{equation}
\left| E \right> = \left[ \begin{array}{c} E_x \\ E_y  \end{array} \right]
	= \left[ \begin{array}{c} \sin \left( \frac{\delta_{CP}}{2} \right) \\
	  e^{i(\frac{\pi}{2} + \delta_{PS})} 
	  \cos \left(\frac{\delta_{CP}}{2} \right) \end{array} \right],
\end{equation} 
where $\delta_{CP}$ and $\delta_{PS}$ are the polarization phase shifts imparted by the CP and PS Pockels cells.  The laser circular polarization following the PS cell is given by 
\begin{equation}
P_\gamma = \sin \delta_{CP} \cos \delta_{PS}.
\end{equation}
Careful attention is paid to reducing the laser$'$s residual linear polarization, since photoemission from strained GaAs can have a significant dependence on this.\cite{prepost}

In the initial setup, with reasonable care for the optics alignment, one finds typical (L-R)/(L+R) charge asymmetries of 1000 parts per million (ppm) and (L-R) position differences at the photocathode of 2 microns.  Three main techniques are then used to reduce these $\beamalr$$'$s:~\cite{humensky}
\begin{enumerate}

\item \underline{Passive setup}
reduces the charge asymmetry to below 100 ppm and the position difference to less than 0.5 microns:
\begin{enumerate}

\item  the ({\it left} or {\it right}) helicity bit information is delayed by one machine rf cycle before transmission to the experiment.  It is rf modulated prior to broadcast and then rf demodulated at the experiment.  This suppresses electronics pickup;
\item  the laser beam is at a waist with rms radius of 1mm at the CP and PS Pockels cells;
\item  The CP Pockels cell is imaged onto the photocathode;
\item  The CP and PS voltages are adjusted to compensate for residual linear polarization in the laser light and reduce the measured charge asymmetry 
\end{enumerate}

\item \underline{Active suppression with feedbacks}
further reduces the charge asymmetry to the 100-ppb level and the position difference at the photocathode to the 100-nanometer level (integrated over the experiment).  Typical {\it left-right} position differences at the E158 target, integrated over a run, are at the 10-nanometer level.  The beam transport and lattice optics provide some demagnification and randomization of the position differences at the source.
\begin{enumerate}

\item  An intensity feedback loop measures the charge asymmetry with electron beam toroids at 1 GeV beam energy.  Control is with the IA Pockels cell in the source laser system.
\item  A position feedback loop measures the position differences with electron beam position monitors (BPMs), also at 1 GeV beam energy.  Control is with the piezo mirror in the source laser system.
\end{enumerate}

\item \underline{Slow reversals}
further suppress effects from charge asymmetries and position differences.
\begin{enumerate}

\item  A half-waveplate to flip the laser helicity can be inserted either after the PS Pockels cell or just before the laser beam enters the accelerator vacuum window in front of the photocathode.  This flips the laser helicity (and hence the $\moller$ physics asymmetry) while leaving certain classes of $\beamalr$$'$s unaffected.  These waveplate flips were done typically every two days in Run I and Run II, and every day in Run III.
\item  The main Linac can be operated at 45.6 GeV or 48.7 GeV, resulting in a flip of the electron beam helicity at the target for a given beam helicity in the Linac.  This flips the $\moller$ physics asymmetry while leaving certain classes of $\beamalr$$'$s unaffected.  One energy flip was done in each of Run I and Run II.  In Run III, the energy was flipped every 2-4 days.
\item   An insertable {\it Asymmetry Inverter (AI)} optics system was used to invert the spatial and angular distributions of the laser beam at the photocathode.  This flips certain classes of $\beamalr$$'$s, while leaving the $\moller$ physics asymmetry unchanged.  In Runs I and II, the {\it AI} optics were flipped once for each energy state.  In Run III they were flipped once in the middle of the run.
\end{enumerate}
\end{enumerate}

\begin{figure}
\centering
\epsfig{file=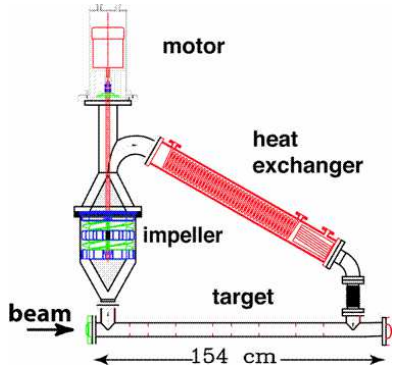,width=2.0in}
\caption{ Schematic of the Liquid Hydrogen Target.}
\label{fig:target}
\end{figure}

\section*{4.  LH$_2$ Target, Spectrometer and Detectors}

The liquid hydrogen target~\cite{targetnim} is 1.57 meters long (0.18 radiation lengths), with a volume of 47 liters and a flow rate of 5 meters/s.  A schematic of it is shown in Figure~\ref{fig:target}.  Eight wire-mesh annular disks in the target cell region introduce turbulence at the 2mm scale (comparable to the beam size) and also induce a transverse velocity component to ensure mixing of the liquid between beam pulses.  The power deposition in the target is 500W.

\begin{figure}
\centering
\epsfig{file=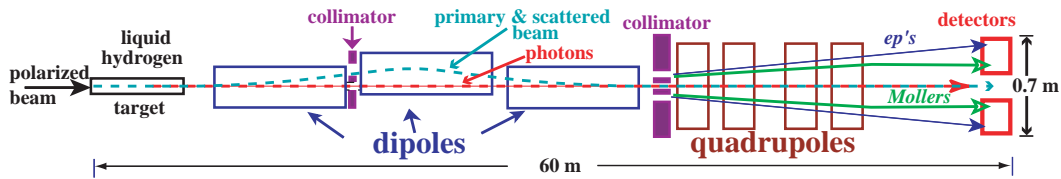,width=5.5in}
\caption{Target, Spectrometer and Detector configuration in End Station A.}
\label{fig:esa}
\end{figure}

\begin{figure}
\centering
\epsfig{file=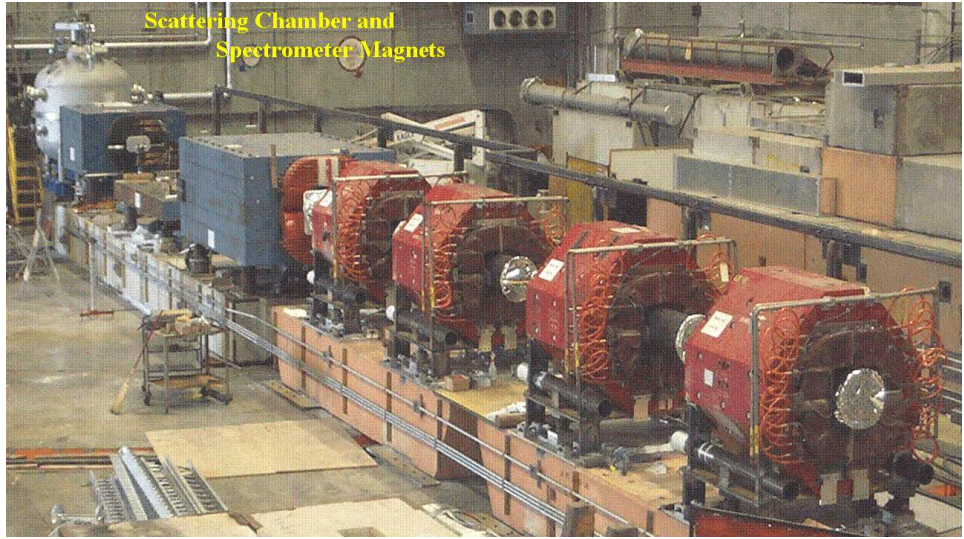,width=5.0in}
\caption{Scattering chamber for the LH$_2$ target and spectrometer magnets (2 of 3 dipoles for the chicane, and 4 quadrupoles) during beamline construction in End Station A.}

\label{fig:spectrometer}
\end{figure}

\begin{figure}
\centering
\epsfig{file=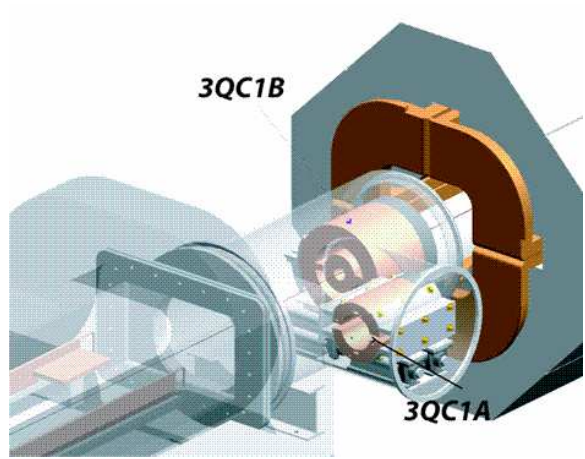,width=3.0in}
\caption{ 3QC1B is the main acceptance collimator located upstream of the first quadrupole.  It has an annular acceptance with inner radius 7.7 cm and outer radius 11.7 cm for $\moller$ (and ep) electrons.  The primary beam and forward photons, produced in the target, pass unimpeded through the spectrometer and detector region to the beam dump.  The central spoke of height 2.6cm prevents synchrotron radiation from the spectrometer dipoles hitting the MOLLER detector.  3QC1A is an insertable collimator used for polarimetry measurements with small acceptance holes to give better separation between the $\moller$ and ep peaks at the detector plane.}
\label{fig:3qc1b}
\end{figure}

A spectrometer consisting of a 3-dipole chicane and 4 quadrupoles (see Figures~\ref{fig:esa} and~\ref{fig:spectrometer}) is used to spatially separate the $\moller$-scattered electrons from the Mott  (electron-proton scattering) background at the detector plane 60 meters downstream of the LH$_2$ target.  Precision collimators are implemented to shield the detector from direct line-of-sight to the target and to provide an angular acceptance of 4.4 mrad $< \theta_{lab} < 7.5$ mrad.   The main acceptance collimator, 3QC1B, is shown in Figure~\ref{fig:3qc1b} and is located just upstream of the first quadrupole.  $\moller$ electrons incident on the MOLLER detector have energies in the range 13-24 GeV.  

\begin{figure}
\centering
\epsfig{file=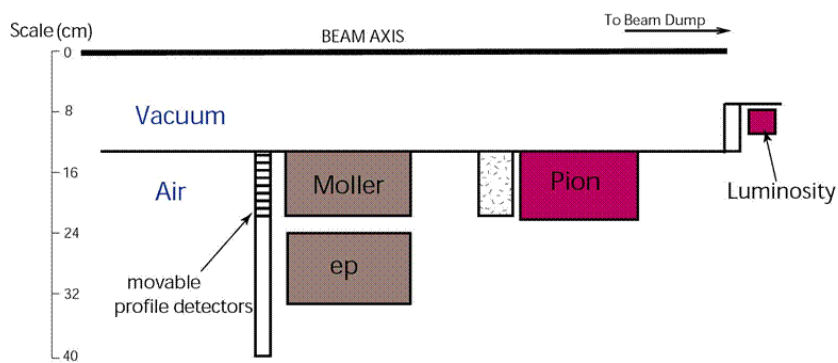,width=4.5in}
\caption{ Detector configuration in End Station A.}
\label{fig:detectors}
\end{figure}

The detector layout is shown in Figure~\ref{fig:detectors}.  There are 4 principle detectors:  MOLLER to observe the $\moller$ signal, ep to observe the Mott background, PION to measure the pion background and LUMI to detect the very forward Mott and $\moller$ electrons with small asymmetry.  Additionally, there is a PROFILE detector to measure the radial electron distributions.  
And there is a  POLARIMETER detector, which is not shown in Figure~\ref{fig:detectors}, but is located on the detector cart just upstream of the MOLLER detector.  The MOLLER, ep, PION and LUMI signals are read into 16-bit custom-built VME ADC's.~\cite{relyea}  The PROFILE and POLARIMETER detector signals are read out by a conventional LeCroy 2249W 11-bit adc.

\begin{figure}
\centering
\epsfig{file=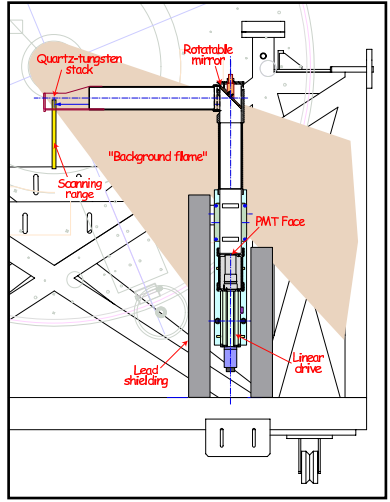,width=2.0in}
\caption{ The polarimeter detector assembly.  A linear drive moves it radially up and down to allow scanning of the $\moller$  and ep peaks.}
\label{fig:polarimeter}
\end{figure}

The POLARIMETER detector assembly is shown in Figure~\ref{fig:polarimeter}.  It consists of a quartz Cherenkov radiator behind a tungsten pre-radiator.  An air light guide directs the Cherenkov light to a PMT.  Beam polarization measurements are performed using polarized $\moller$ scattering with a polarized supermendur foil target (20, 50 and 100 micron thickness foils were used) and using the same spectrometer as for the main experiment.  The polarized target foils can be inserted into the beam just upstream of the LH$_2$ target station.  The foils are pitched with respect to the beam by 20$^\circ$ and are polarized along the foil direction by a magnetic field of 90 gauss.  The beam polarization is measured to be $(85 \pm 5)\%$ for Runs I and Run II, where the $5\%$ systematic error is dominated by the uncertainty in the target foil polarization ($3\%$) and the background subtractions ($3\%$).  Polarization measurements were performed once every 2 days by taking 10 minutes of data with the foil in and another 2 minutes of {\it foil out} data.  The entire polarimetery measurement procedure, including setup and recovery to production data, takes about 45 minutes. 

\begin{figure}
\centering
\epsfig{file=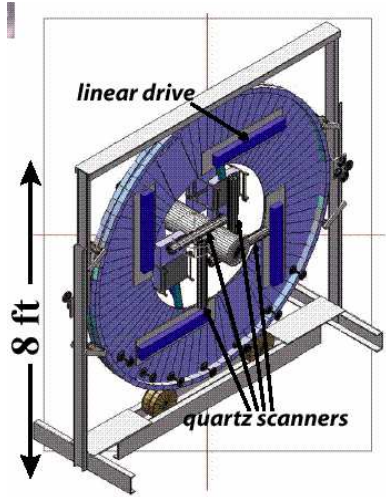,width=2.0in}
\caption{ Profile Detector schematic.  4 scanners mounted on the wheel can each move radially; the wheel itself can rotate the scanners to different azimuthal angles.}
\label{fig:profile}
\end{figure}

\begin{figure}
\centering
\epsfig{file=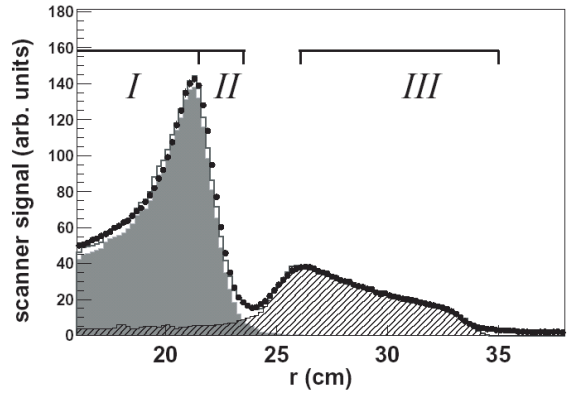,width=3.0in}
\caption{ A profile scan by the PROFILE detector.  The location of the MOLLER detector, regions I and II, and the ep detector, region III, are indicated .  Data are shown by closed circles.  The Monte Carlo result is given by the open histogram, with contributions from $\moller$ (shaded) and ep (hatched) shown separately.}
\label{fig:profilescan}
\end{figure}

\begin{figure}
\centering
\epsfig{file=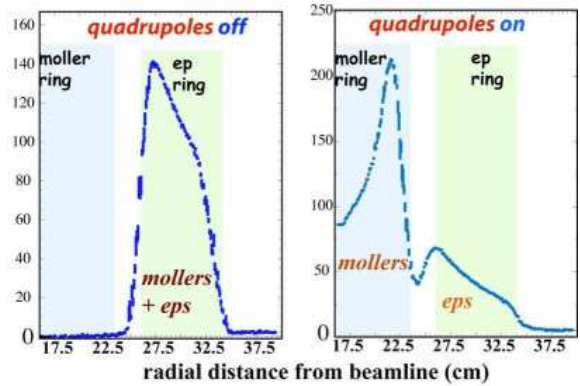,width=3.0in}
\caption{Profile scan with the PROFILE detector, comparing {\it quads on} and {\it quads off}.}
\label{fig:qonoff}
\end{figure}

The PROFILE detector is located just upstream of the MOLLER and ep detectors and is shown in Figure~\ref{fig:profile}.  Many detailed, dedicated profile scans were taken in each of the 3 physics runs to assist accurate determinations of different sources of backgrounds and to verify the geometry for the Monte Carlo simulation of the experiment.  A typical profile scan is shown in Figure~\ref{fig:profilescan}, with the relative contributions from $\moller$ and ep contributions indicated (as determined from Monte Carlo).  The radial acceptance of the MOLLER (Regions I and II) and ep (Region III) detectors is also indicated.  The many detailed profile scans performed included the following configurations:
\begin{itemize}
\item spectrometer quads on and off as well as $\pm 5\%$ variations of nominal strength
\item LH$_2$ target in and out
\item (remotely insertable) collimators in and out
\item 45.6 and 48.7 beam energies
\end{itemize}
Figure~\ref{fig:qonoff} shows one example of a profile scan comparing {\it quads on} with {\it quads off}, illustrating the suppression of the $\moller$ peak with {\it quads off}.

\begin{figure}
\centering
\epsfig{file=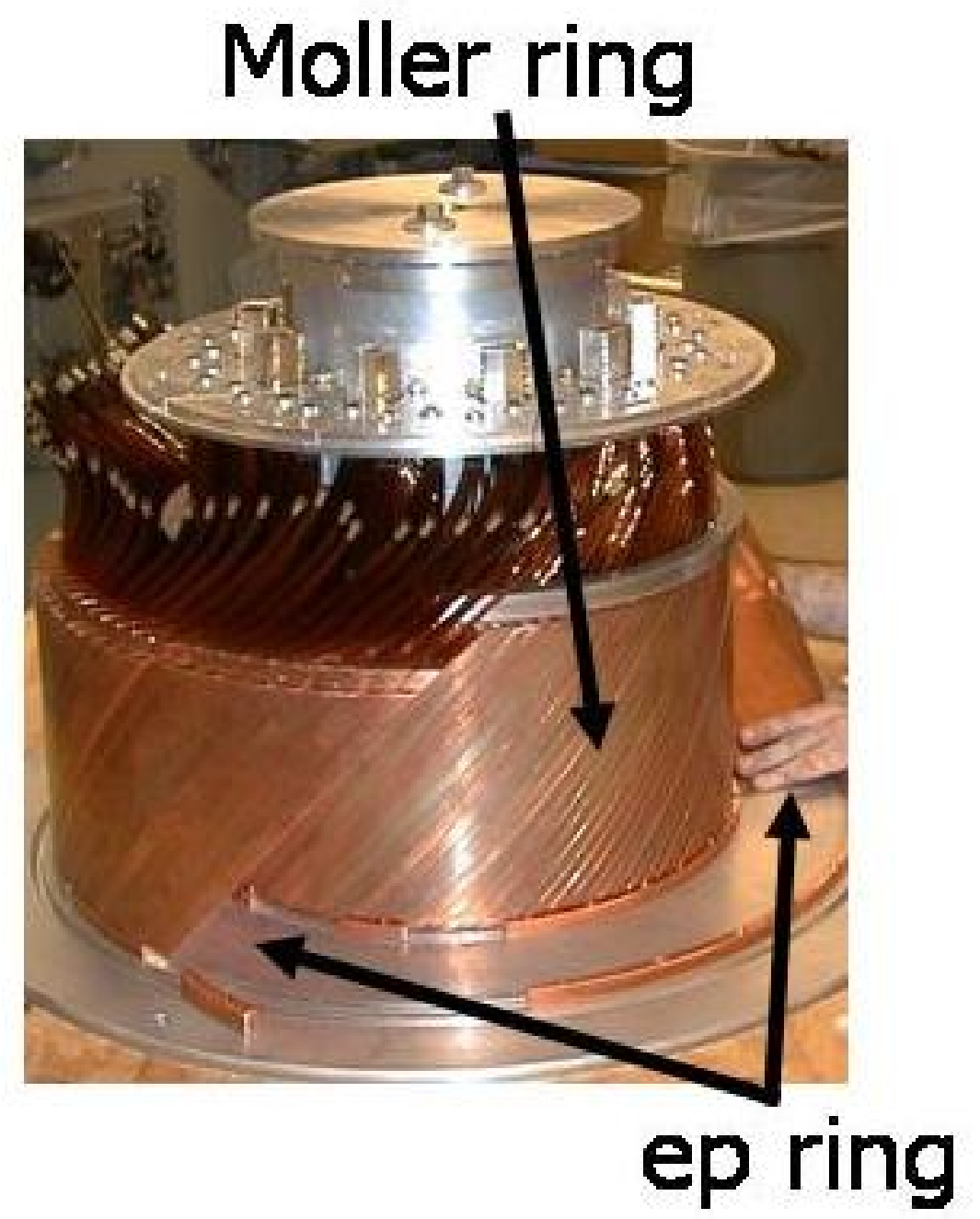,width=2.5in}
\caption{MOLLER-ep Detector during assembly.}
\label{fig:mollerdetector}
\end{figure}

The MOLLER and ep detectors are sampling calorimeters with fused silica fibers oriented at the Cherenkov angle for the active medium and copper plates for the radiator.  The Cherenkov light from the fibers in these detectors is directed to 60 shielded PMTs via air light guides, providing both azimuthal and radial segmentation.  A photo of these detectors during assembly is shown in Figure~\ref{fig:mollerdetector}.  The radial segmentation of the MOLLER and ep detectors is shown in Figure~\ref{fig:profilescan}.  Approximately 20 million electrons with an average energy of 17 GeV are incident on the MOLLER detector each spill.

\begin{figure}
\centering
\epsfig{file=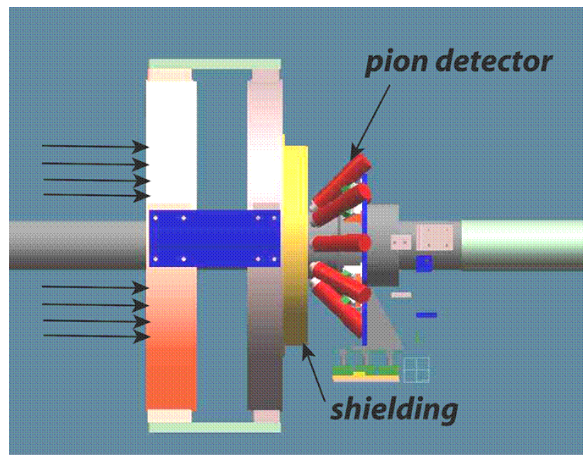,width=3.0in}
\caption{ Schematic of the PION detector, located behind a thick lead radiator behind the MOLLER detector.}
\label{fig:pion}
\end{figure}

The PION detector, shown in Figure~\ref{fig:pion}, is a quartz bar radiator with PMT readout.  Its measurements determine the pion contamination to the MOLLER signal to be $(0.1 \pm 0.1)\%$ with an asymmetry of 1 ppm.  This results in a correction to the measured MOLLER asymmetry of $(1 \pm 1)$ ppb.  

\begin{figure}
\centering
\epsfig{file=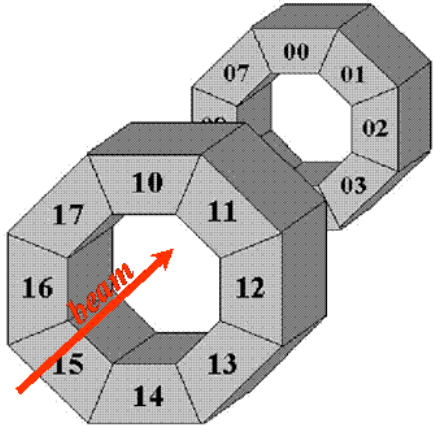,width=2.5in}
\caption{ Schematic of the LUMI Detector.}
\label{fig:lumi}
\end{figure}

LUMI is a segmented ion chamber with aluminum pre-radiator.  It has 2 separate longitudinal rings and is shown in Figure~\ref{fig:lumi}.  LUMI sees a flux per spill of $\approx 350$ million electrons, with a roughly equal mix of $\moller$ and Mott electrons.  The mean electron energy in LUMI is 40 GeV and the expected physics asymmetry is (-15 $\pm$ 5) ppb.  LUMI performs an important "null" cross check for the experiment.  It observes lower ($\approx 1.5$ mrad) angle electron scattering than the main MOLLER detector ($\approx 6.0$ mrad) and is therefore more sensitive to beam fluctuations and $\beamalr$$'$s.  Additionally, it should have the same sensitivity as the MOLLER detector to LH$_2$ target density asymmetries that could arise from beam spotsize asymmetries.~\cite{targetnim} In Run I, its measured asymmetry was $(-16 \pm 16)$ ppb; and in Run II, it measured $(-14 \pm 17)$ ppb.  Both these results are consistent with expectations and help set limits on false beam asymmetries.

\vspace{8.5mm}
\section*{5.  Analysis}
Every 33 milliseconds 4 beam spills are recorded in the detectors and beam monitors.  The helicity of each spill is chosen in a pseudo-random sequence of pulse quadruplets, $R_1R_2\overline{R_1}\overline{R_2},R_3R_4\overline{R_3}\overline{R_4}...$  The 
helicity of the first two pulses are chosen pseudo-randomly, while the 
next two states are complements of the first two.  Then two more pseudo-random
helicity states are chosen and so on.  Each quadruplet of beam spills consists of two (L,R) pulse pairs--one on each of the 60Hz beam timeslots.  This fast random switching of the beam helicity suppresses $\beamalr$$'$s and suppresses effects from target density fluctuations.  The quadruplet structure also suppresses jitter from timeslot effects.  

Detector signals are first normalized to toroid measurements of the beam charge and then detector asymmetries, $A^p_{exp}$, are formed from each (L,R) pair of pulses, p,:
\begin{equation}
A^p_{exp}=\frac{\frac{{\rm Det}_R}{Q_R}-\frac{{\rm Det}_L}{Q_L}}{\frac{{\rm Det}_R}{Q_R}+\frac{{\rm Det}_L}{Q_L}} 
\end{equation}
These detector asymmetries are corrected for differences in the {\it right-left} beam properties as measured by the BPMs and toroids,
\begin{equation}
A_{raw}^{p}=A_{exp}^{p}- \sum_{i=E,x,y,\theta_x,\theta_y,Q}
 a_i \Delta x_i
\end{equation}
where $\Delta x_i$ are the {\it right-left} differences in energy, position, angle and charge.  The coefficients $a_i$ are determined by two independent methods.  The first uses an unbinned linear least squares linear regression.  The second technique utilizes beam dithering measurements.  We dither 5 independent beam "coils" at the end of the Linac (these coils are indicated in Figure~\ref{fig:beammonitors}).  The energy dither, with an amplitude of $\pm 50$ MeV, uses pulsed phase shifters for the Sector 27 and Sector 28 sub-boosters.  (The Linac has 30 sub-boosters, one for each of the 30 Linac sectors.)  Four pulsed correctors are used to dither x, $\theta_x$, y, and $\theta_y$.  The response of the energy, position and angle BPMs to these dither coils as well as the response of the detector signals to these coils is measured, allowing the coefficients $a_i$ to be determined.  Approximately 5$\%$ of the beam pulses were dithered, with the dithering amplitudes chosen to be about 2-3 times the beam jitter.  The regression and dithering techniques yield consistent results for the physics asymmetries measured, and we choose to use the statistically more powerful regression technique for the results quoted.

\begin{figure}
\centering
\epsfig{file=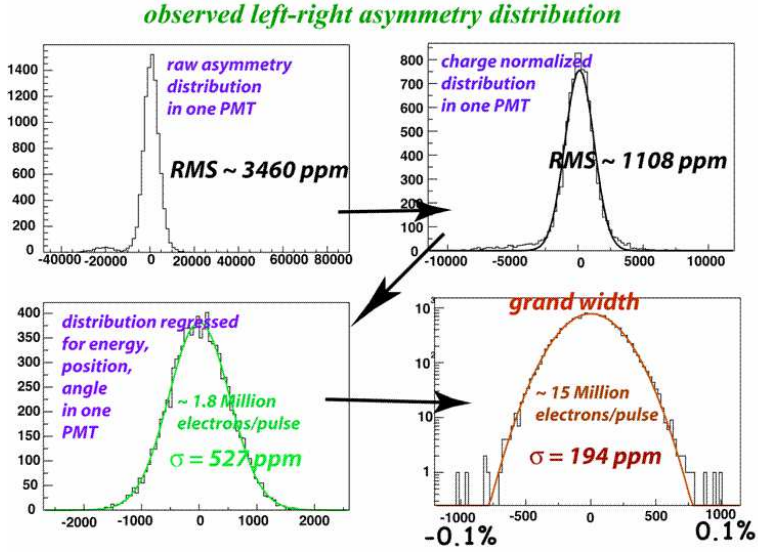,width=4.0in}
\caption{ Regression analysis for the MOLLER detector.}

\label{fig:regression}
\end{figure}

The effects of the regression procedure for the MOLLER detector are illustrated in Figure~\ref{fig:regression}.  The first frame of this figure shows the raw asymmetry distribution in a single PMT (which covers a small azimuthal range).  The distribution has an rms of 3460 ppm, consistent with the beam intensity jitter of 0.5$\%$ per pulse.  Normalizing the PMT signal to the toroids reduces the rms to 1108 ppm.  Regression against the energy, position and angle BPMs reduces the rms further to 527 ppm.  Summing all the MOLLER PMTs (integrating over azimuth and radius) then results in an rms width of 194 ppm.

After performing the beam asymmetry corrections to the pulse-pair asymmetry, we sum over all pulse pairs,
\begin{equation}
A_{raw}=\sum_{p}A_{raw}^p.
\end{equation}
At this stage we perform two important checks on the beam asymmetry corrections that have been applied.  First, we check that independent beam monitors give consistent results for the beam corrections to $A_{raw}$ for the MOLLER detector.  For Run I, this consistency is summarized in Table~\ref{tab:monitoragree}.  (Similar results are obtained for Run II.)  The consistency is excellent and the uncertainties reflect the beam monitor resolutions.  These resolutions make a small contribution to the total statistical error for the $\moller$ $\apv$ measurement.  Second, we check that regression and dithering give consistent results.  

\begin{table} [tbp]
\caption{Agreement of independent beam monitors for the left-right beam asymmetries, $\beamalr$$'$s, for Run I.  (The beam monitor agreements in Run II are similarly good.)} 
\begin{center}
\begin{tabular}{|c|c|c|c|}
\hline
{\bf Beam}	  & {\bf  Beam}      & {\bf Monitor}      & {\bf MOLLER Correction}	\\
{\bf Parameter} & {\bf Monitors} & {\bf Agreement} & {\bf Agreement} \\
\hline
E 		& BPMs 12X, 24X	& $(0.00 \pm 0.24)$ keV	& $(0.1 \pm 1.3)$ ppb 	\\
X		& BPMs 41X, 42X	& $(1.0 \pm 0.6)$ nm		& $(0.9 \pm 0.5)$ ppb		\\
Y		& BPMs 41Y, 42Y	&  $(0.0 \pm 1.0)$ nm	& $(0.0 \pm 0.2)$ ppb	\\
$\theta_x$	& BPMs 31X, 32X 	&  $(-2.8 \pm 2.0)$ nm	& $(-2.4 \pm 1.9)$ ppb	\\
$\theta_y$	& BPMs 31Y, 32Y	&  $(0.9 \pm 1.0)$ ppb	& $(0.7 \pm 0.8)$ ppb		 \\
Q		& Toroids 2a, 3a	&  $(-2.9 \pm 5.3)$ ppb	& $(-2.9 \pm 5.3)$ ppb	\\
\hline
\end{tabular}
\end{center}
\label{tab:monitoragree}
\end{table}

Finally, we correct $A_{raw}$ for backgrounds and dilutions and normalize it to the beam polarization and detector linearity, to arrive at the physics asymmetry, 
\begin{equation}
A_{PV} = \frac{1}{P_b \cdot \epsilon} \cdot \frac{A_{raw}-\sum_{bkg}f_{bkg}A_{bkg}}{(1-\sum_{bkg}f_{bkg})}.
\label{eq:apv}
\end{equation}
$P_b$ is the beam polarization, measured by the polarimeter to be $P_b = (0.85 \pm 0.05)$ for Runs I and II; $\epsilon$ is the linearity, which for the MOLLER calorimeter is determined to be $\epsilon = 0.99 \pm 0.01$; and the $f_{bkg}$ and $A_{bkg}$ give the dilution and asymmetry contributions of each background source (these are summarized for the Run I $\moller$ asymmetry analysis in Table~\ref{tab:systematics}).
\section*{6.  Results}
We present results using data collected in Runs I and II, which comprise 50$\% $ of the total data sample.  The Run I results for the weak mixing angle determined from $\apv$ in $\moller$ scattering have been published,\cite{run1prl} and Run II results for this are preliminary.   We also present preliminary results for the transverse e-e- asymmetry measured during Run I, and for the longitudinal ep asymmetry also using Run I data.

\subsection*{6.1 Transverse e-e Asymmetry}
\begin{figure}
\centering
\epsfig{file=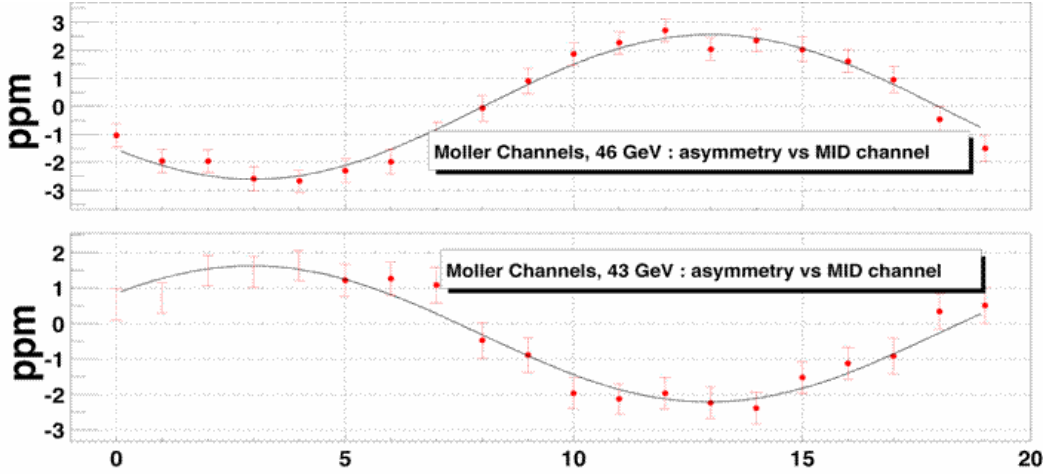,width=5.5in} 
\caption{Observed raw asymmetry in the MOLLER MID detector ring (in Region I) for transverse beam polarization incident on the LH$_2$ target.  The MID ring has 20 azimuthal segments.}
\label{fig:eetransverse}
\end{figure}

First we present a measurement of the transverse e-e asymmetry observed in special runs taken during Run I.  The raw MOLLER asymmetry, $A_{transverse}^{ee}$, in Region I is plotted in Figure~\ref{fig:eetransverse} versus azimuth for beam energies of 43 GeV and 46 GeV.  The horizontal beam polarization yields an $\approx 2.5$ ppm
up-down asymmetry, which flips sign as expected between the two beam energies.  This asymmetry is due to a 2-photon exchange QED process and is consistent with theoretical expectations.~\cite{dixon}  The analysis of this data is continuing, and we have additional data for this taken during Run II and Run III.   We hope to achieve 2-3$\%$ statistical error on the measurement.  The theoretical prediction, including radiative corrections, has less than 1$\%$ uncertainty,\cite{dixon} but depends on the energy and angular acceptance of the E158 spectrometer and detector.  If acceptance uncertainties allow for a prediction at the $\approx 2\%$ level, we can use this data to cross check our normalization factors for beam polarization, linearity, and dilution factors.

\subsection*{6.2 ep Asymmetry}
\begin{figure}
\centering
\epsfig{file=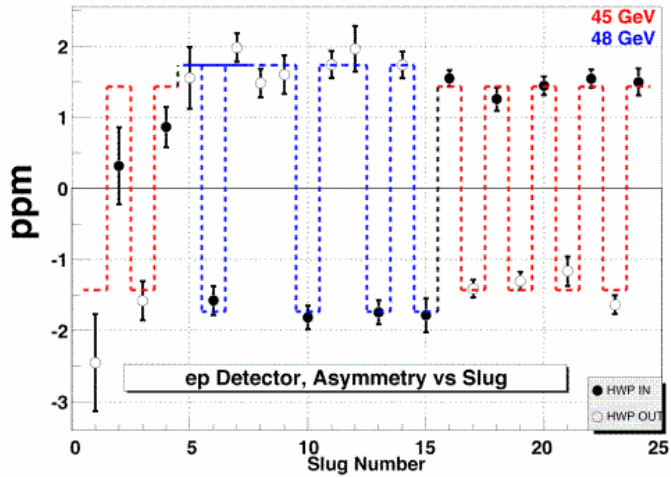,width=3.5in} 
\caption{ Observed asymmetry, $A_{raw}^{ep}$ in the "ep" detector.  The dashed line indicates the expected asymmetry, taking into account the beam energy and the source half-waveplate status.  A "slug" refers to a data sample ($\approx$2 days) with a consistent half-waveplate and energy configuration.}
\label{fig:epasym}
\end{figure}

We measure the asymmetry, $\apv^{ep}$, in Region III of the MOLLER-ep detector which should be dominated by the inelastic ep asymmetry.  The preliminary results for $A_{raw}^{ep}$, uncorrected for sign flips from the two half-waveplate and two energy states, using Run I data are shown in Figure~\ref{fig:epasym}.  We measure
\begin{eqnarray}
A_{Raw}^{ep}(45 GeV) & = & (-1.36 \pm 0.05) \mbox{ppm (stat. only)} \\
A_{Raw}^{ep}(48 GeV) & = & (-1.70 \pm 0.08) \mbox{ppm (stat. only)} \\
\frac{\apv^{ep}(48 GeV)}{\apv^{ep}(45 GeV)} & = & 1.25 \pm 0.08 \mbox{ (stat) $\pm$ 0.03 (syst)}
\end{eqnarray}
The size of asymmetry is consistent with $-10^{-4} \cdot Q^2$, which is the expectation for deep inelastic scattering.  The ratio of asymmetries is consistent with the kinematic variation of $Q^2$ between the two energies. 

\subsection*{6.3 $\moller$ Asymmetry}
The primary goal for E158 is a precise measurement of the $\moller$ asymmetry.  The MOLLER detector Region I signals are used for this analysis.  A summary of corrections to $A_{raw}^{\moller}$ for beam asymmetries, residual transverse polarization and background contributions is given in Table~\ref{tab:systematics} for the Run I data.  

The first-order beam corrections total -41 ppb as determined from regressing against the beam monitors.  A number of studies are performed to examine effects from higher-order beam asymmetries, such as beam spotsize asymmetries or energy spread asymmetries.  The wire array directly measures spotsize asymmetries and we measure the correlation dependence of MOLLER detector asymmetries with these spotsize asymmetries.  In Run I we find $\Delta A_{spotsize}=0.1 \pm 0.5$ ppb.  Limits on other higher-order beam asymmetry effects are set by studying detector MONITORS that have much larger sensitivities (by a factor 10 or more) to the first-order beam parameters (energy, position etc.).  (In particular, we construct TIMESLOT MONITORS that take advantage of sometimes large beam asymmetries on individual timeslots.  The beam feedbacks keep the average of the beam asymmetries on the two 60Hz timeslots small, but beam asymmetries on individual timeslots can sometimes be large.  We then observe how well the MONITOR asymmetries regress in the presence of large beam asymmetries.)  These MONITORS include the Region II MOLLER detector (sensitive to energy), the LUMI detector (sensitive to energy and angles) and detector DIPOLES (Region II DIPOLES are sensitive to beam position at the LH$_2$ target, and LUMI DIPOLES are sensitive to beam angles at the target).  XDIPOLES and YDIPOLES can be formed for each detector using the azimuthal segmentation and using an appropriate $\cos \phi$ or $\sin \phi$ weighting scheme for the asymmetry contribution from each segment.  The DIPOLES have much greater sensitivity to beam first-order moments than the MONOPOLES ($\approx$equal weighting for each azimuthal segment) used for the $A_{PV}$ measurements.  The relative sensitivities to first-order beam moments of the MONITORS compared to the $A_{PV}$ MONOPOLES can be directly measured and the relative sensitivities to higher-order beam moments should be even greater.  These studies lead to $\Delta A_{\rm higher-order}=0 \pm 10$ ppb.  A further check on the beam asymmetry corrections is provided by the agreement of the {\it regression} and {\it dithering} analyses.  In Run I the difference in corrections determined by these two methods is $(3.1 \pm 11.7)$ ppb, and for Run II it is $(3.2 \pm 3.8)$ ppb.

There is a small correction to $A_{PV}$ due to residual transverse polarization of the beam (resulting from the large e-e transverse physics asymmetry discussed above and the beam energy not being tuned exactly), which we determine to be $(-8 \pm 3)$ ppb from measurements of the MOLLER YDIPOLE.  

The background corrections that appear in Equation~\ref{eq:apv} are summarized in Table~\ref{tab:systematics} for the Run I data.  The largest correction of -26 ppb is due to electrons from inelastic ep scattering and electrons from photon conversions to $e^+e^-$ pairs.  The measured asymmetry in the Region III ep detector was used, together with Monte Carlo simulations, to estimate this as well as the smaller ep elastic correction of -8 ppb.  The background correction from the pion background is determined from a direct measurement of the pion flux and asymmetry in the PION detector and a Monte Carlo simulation of the pion energy deposition in the MOLLER detector.  The background corrections for high energy photons, synchrotron photons and neutrons were determined from a series of dedicated studies using {\it quads off} data (high energy photon background), {\it target out} data (synchrotron photon background) and {\it blinded phototube} data (neutrons and neutral hadrons backgrounds).  (The asymmetry in the synchrotron photon background is estimated by using the observed MOLLER XDIPOLE measurements, together with the measured $A_{transverse}^{ee}$, to constrain the vertical component of the beam polarization, and a Monte Carlo simulation of the MOLLER detector$'$s energy response.)  To accurately assess the level of these backgrounds required special run configurations to test the detector linearity.  This was done with a series of runs including the following configurations:  different PMT gains, LH$_2$ target, 2g and 8g carbon targets, 100-micron polarized foil target and no target.  
 
\begin{table} [tbp]
\caption{$\apv$ Corrections, $\Delta A$, and dilution factors, $f$, for the $\moller$ asymmetry analysis in Run I.} 
\begin{center}
\begin{tabular}{|c|c|c|}
\hline
{\bf Source}				& {\bf $\Delta A$ (ppb)} & {\bf $f$}	\\
\hline
Beam (first order) 			& $-41 \pm 3$		& 	\\
Beam (higher order)			& $0 \pm 10$		& 	\\
Transverse polarization		& $-8 \pm 3$		&	\\
$e^-+p \rar e^-+p(+\gamma)$ 	& $-8 \pm 2$ 		&  $0.064 \pm 0.007$ \\
$e^-(\gamma)+p \rar e^-$ +X		& $-26 \pm 6$		&  $0.011 \pm 0.003$ \\
Pions					& $1 \pm 1$		&  $0.001 \pm 0.001$ \\
High energy photons			& $3 \pm 3$		&  $0.004 \pm 0.002$ \\
Synchrotron photons			& $0 \pm 5$		&  $0.002 \pm 0.001$ \\
Neutrons				& $-5 \pm 3$		&  $0.003 \pm 0.001$ \\
\hline
\end{tabular}
\end{center}
\label{tab:systematics}
\end{table}

The corrections, $\Delta A$, and dilution factors, $f$, for Run II data are similar to those in Table~\ref{tab:systematics} with a few exceptions.  The values for beam (higher order), pions, high energy photons and neutrons are identical.  Beam (first order) corrections are $(-19 \pm 3)$ ppb.  The transverse polarization correction is $(-5 \pm 3)$ ppb.  The synchrotron photon correction is $(0 \pm 2)$ ppb.  The ep corrections are reduced and total $(-29 \pm 4)$ ppb.  The ep dilution factor is $(0.062 \pm 0.007)$.  The improvement here is due to more optimized quadrupole spectrometer settings and to an additional collimator (CM8) that was installed between Run I and Run II to absorb the ep flux incident on the ep detector, which reduced shower leakage from the ep detector to the MOLLER detector.

\begin{figure}
\centering
\epsfig{file=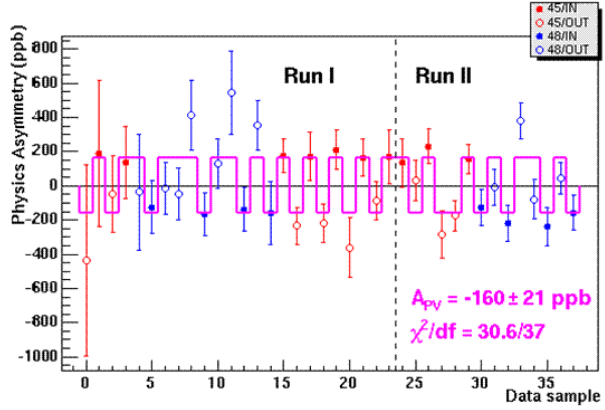,width=3.5in} 
\caption{ Observed $\moller$ physics asymmetry in Runs I and II.  The solid line indicates the expectation for the asymmetry.  Each data sample corresponds to approximately 2 days of running time.}
\label{fig:mollerapv}
\end{figure}

\begin{figure}
\centering
\epsfig{file=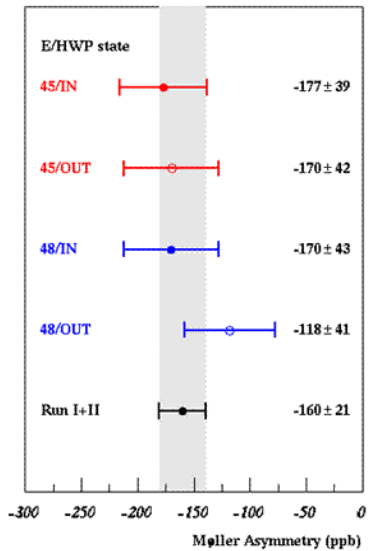,width=2.2in} 
\caption{ $\moller$ physics asymmetry, separately for each of the energy and half-waveplate configurations.}
\label{fig:apvlep}
\end{figure}

We use Equation~\ref{eq:apv} to correct $A_{raw}^{MOLLER}$ for the beam polarization, detector linearity, and the transverse polarization and background contributions to arrive at the $\moller$ physics asymmetry, $\apvmoller$.  (The beam first order corrections are already included in $A_{raw}^{MOLLER}$.)  Our published result using Run I data only gives
\begin{equation}
\apv \mbox{ ($e^-e^-$ at Q$^2$= 0.026 GeV$^2$) =-175 $\pm$ 30 (stat.) $\pm$ 20 (syst.) ppb}
\end{equation}
Including our preliminary results using Run II data, we find
\begin{equation}
\apv \mbox{ ($e^-e^-$ at Q$^2$= 0.026 GeV$^2$) =-160 $\pm$ 21 (stat.) $\pm$ 17 (syst.) ppb}
\end{equation}
This $\moller$ asymmetry is plotted versus data sample in Figure~\ref{fig:mollerapv}.  The sign flips due to the energy state and half-waveplate state are not included so as to illustrate that the measured asymmetry has the expected sign flips with respect to energy and half-waveplate configurations.  The consistency of $\apvmoller$ for each of the 4 energy and half-waveplate configurations is shown in Figure~\ref{fig:apvlep}.

\section*{7.  Weak Mixing Angle}

\begin{figure}
\centering
\epsfig{file=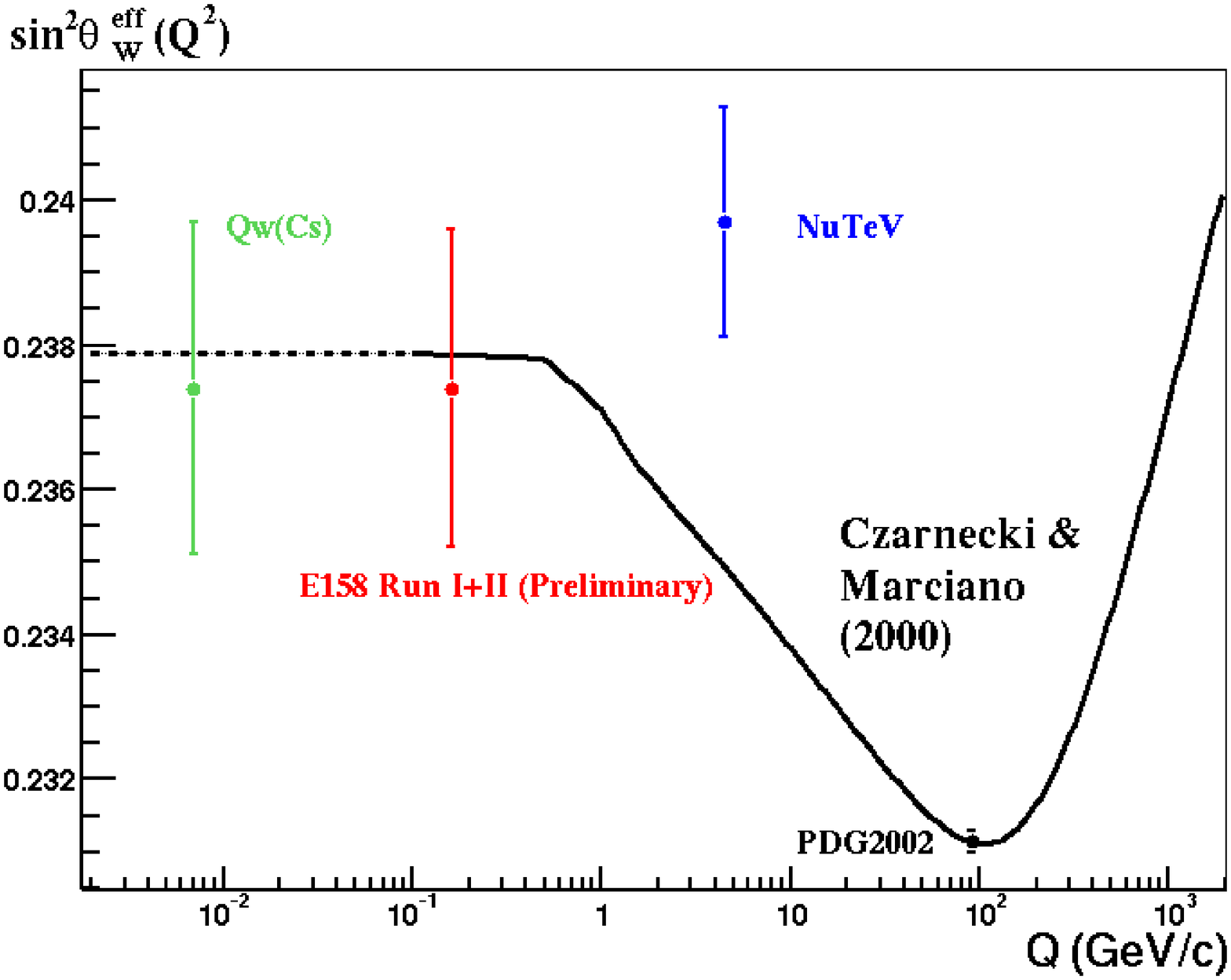,width=3.5in} 
\caption{Measurements of the Weak Mixing Angle as a function of $Q^2$, together with a theoretical prediction (solid line curve) from Czarnecki and Marciano.\cite{marciano3}}
\label{fig:marcianoplot}
\end{figure}

The tree level expression relating $\apv$ to the weak mixing angle was given in Equation~\ref{eq:apvtree}.  Including radiative corrections, this relation becomes
\begin{eqnarray}
A_{PV} & = & \frac{G_FQ^2}{\sqrt{2} \pi \alpha} \cdot \frac{1-y}{1+y^4+(1-y)^4} \cdot F_{brem} \cdot (1-4 \thetaweff) \\
	 & = & AP \cdot F_{brem} \cdot (1-4 \thetaweff)
\end{eqnarray}
The analyzing power, $AP$, depends on kinematics and the experimental geometry and is determined from a detailed Monte Carlo of the experiment.  Its uncertainty is estimated to be 1.7$\%$.  $F_{brem} = (1.01 \pm 0.01)$ is a correction for initial and final state radiation.~\cite{zykunov}  $\thetaweff$ is derived from an effective coupling constant, $g_{ee}^{eff}$, for the Zee coupling.  Loop and vertex electroweak corrections are absorbed into $g_{ee}^{eff}$.
We use the $\msbar$ scheme to evaluate $\thetaweff$ and we find this {\it Preliminary} result, using both Run I and Run II data,
\begin{equation}
\sintwQ2 = 0.2379 \pm 0.0016 {\rm (stat.)} \pm 0.0013 {\rm (syst.)} \end{equation}
This result, together with other experimental measurements, is presented in Figure~\ref{fig:marcianoplot}.  The SLD and LEP measurements were discussed earlier in Section 1.  The NuTeV result~\cite{nutev} is determined from measurements of the relative rates for neutral current and charged current interactions in neutrino-nucleon scattering.  The APV result~\cite{cesium} is a measurement from atomic parity violation; parity-violating admixtures of atomic states result in an asymmetry in the excitation rates between atomic levels that are measured with a polarized laser beam.

\begin{figure}
\centering
\epsfig{file=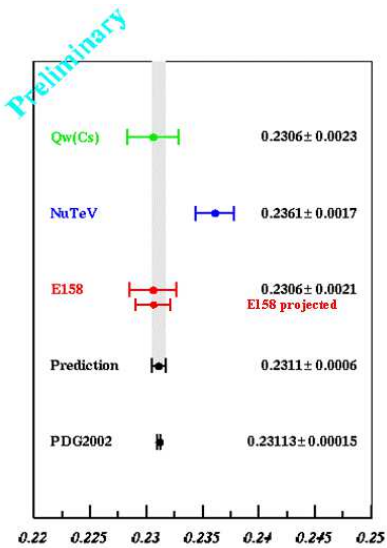,width=2.2in} 
\caption{Summary of Weak Mixing Angle measurements, evolved to the Z-pole.  Also shown is the expectation for the precision E158 should achieve with its full data sample.}
\label{fig:weakanglesummary}
\end{figure}

To more directly compare results among the different experiments, we evolve all measurements to the Z-pole.  The {\it Preliminary} E158 result (using Run I and Run II data) is
\begin{equation}
\sintwMZ = 0.2306 \pm 0.0021.
\end{equation}
Our result is plotted, together with the other experiments, in Figure~\ref{fig:weakanglesummary}.  Also shown is the projected E158 result, when analysis is complete using the full data sample, including Run III.

\section*{8.  Conclusions}
E158 has made the first observation of parity violation in $\moller$ scattering.  We report a {\it Preliminary} measurement from our Run I and Run II data, 
\begin{equation}
\apv \mbox{ ($e^-e^-$ at Q$^2$= 0.026 GeV$^2$) =-160 $\pm$ 21 (stat.) $\pm$ 17 (syst.) ppb}
\end{equation}
with a significance of 6.3$\sigma$ for observing parity violation.  E158 has also made the first observation of a single spin transverse asymmetry in $\moller$ scattering.  This transverse asymmetry is due to a 2-photon exchange QED effect, and may prove useful for cross checking the normalization factors for beam polarization, linearity and dilutions. 
In the context of the Standard Model, the {\it Preliminary} $\apvmoller$ result determines the weak mixing angle to be, 
\begin{equation}
\sintwMZ = 0.2306 \pm 0.0021.
\end{equation}
When the full data set, including Run III, is analyzed we expect to achieve a total error on $\sintwMZ$ below $\pm 0.0015$.

\end{document}